\documentclass[pdflatex,sn-mathphys-num,table,xcdraw]{sn-jnl}

\usepackage{graphicx}%
\usepackage{multirow}%
\usepackage{amsmath,amssymb,amsfonts}%
\usepackage{amsthm}%
\usepackage{mathrsfs}%
\usepackage[title]{appendix}%
\usepackage{subcaption}
\usepackage{xcolor}%
\usepackage{textcomp}%
\usepackage{manyfoot}%
\usepackage{booktabs}%
\usepackage{algorithm}%
\usepackage{algorithmicx}%
\usepackage{algpseudocode}%
\usepackage{listings}%
\usepackage{array}
\usepackage{tabularx}
\usepackage{cleveref}
\usepackage{todonotes}
\usepackage{comment}

\theoremstyle{thmstyleone}%
\theoremstyle{thmstyletwo}%

\theoremstyle{thmstylethree}%

\theoremstyle{definition}

\newtheorem*{proof*}{Proof}

\Crefname{lemma}{Lemma}{Lemmas}
\Crefname{theorem}{Theorem}{Theorems}

\newcommand{\tildeg}{\tilde{g}}

\newcommand{\spher}{\Psi^{-1}}

\begin{document}

\title[Article Title]{Synergistic interplay of morphology and metabolic activity rule response to CAR-T cells in B-cell lymphomas}


\author[1,4]{\fnm{Yifan} \sur{Chen}}
\equalcont{These authors contributed equally to this work.}
\author*[1,2,3]{\fnm{Soukaina} \sur{Sabir}}\email{souky.sabir@gmail.com}
\equalcont{These authors contributed equally to this work.}

\author[4]{\fnm{Christina} \sur{Kuttler}}
\author[1,2,3]{\fnm{Juan} \sur{Belmonte-Beitia}}
\author[5]{\fnm{Alvaro} \sur{Mártínez-Rubio}}
\author[6]{\fnm{Lourdes} \sur{Martín-Martín}}
\author[7]{\fnm{Lucía} \sur{López-Corral}}
\author[7]{\fnm{Alejandro} \sur{Martín-Sancho}}
\author[7]{\fnm{J. Cristobal} \sur{Cañadas Salazar}}
\author[7]{\fnm{Carlos} \sur{Montes-Fuentes}}
\author[7]{\fnm{M. Pilar} \sur{Tamayo-Alonso}}
\author[8]{\fnm{Angel} \sur{ Cedillo}}
\author[8]{\fnm{Pascual} \sur{Balsalobre}}
\author[9]{\fnm{Pere} \sur{Barba}}
\author[10,11]{\fnm{Antonio} \sur{Pérez-Martínez}}
\author*[1,2,3]{\fnm{Víctor M.} \sur{Pérez-García}}\email{victor.perezgarcia@uclm.es}

\affil[1]{\orgdiv{Mathematical Oncology Laboratory (MOLAB), Instituto de Matemática Aplicada a la Ciencia y la Ingeniería}, \orgname{University of Castilla-La Mancha}, \orgaddress{\country{Spain}}}

\affil[2]{\orgdiv{Department of Mathematics, Escuela Técnica Superior de Ingeniería Industrial}, \orgname{University of Castilla-La Mancha}, \orgaddress{\country{Spain}}}

\affil[3]{\orgdiv{Laboratorio de Oncología Matemática, Instituto de Investigación Sanitaria
de Castilla-La Mancha (IDISCAM)}, \orgaddress{\country{Spain}}}

\affil[4]{\orgdiv{Department of Mathematics}, \orgname{TUM School of Computation, Information and Technology, Technical University of Munich}, \orgaddress{\street{Boltzmannstr. 3}, \city{Garching}, \postcode{85747}, \country{Germany}}}

\affil[5]{ \orgname{Institut Curie}, \orgaddress{\city{Paris}, \country{France}}}
\affil[6]{ \orgname{Universidad de Salamanca}, \orgaddress{\city{Salamanca}, \country{Spain}}}
\affil[7]{ \orgname{Hospital Universitario de Salamanca, Instituto de Investigación Biomédica de Salamanca (IBSAL)}, \orgaddress{\city{Salamanca}, \country{Spain}}}
\affil[8]{ \orgname{Grupo Español de Trasplante Hematopoyético y Terapia Celular}, \orgaddress{\city{Madrid}, \country{Spain}}}
\affil[9]{ \orgname{Hospital Universitari Vall d'Hebron}, \orgaddress{\city{Barcelona}, \country{Spain}}}
\affil[10]{ \orgname{University Hospital La Paz}, \orgaddress{\city{Madrid}, \country{Spain}}}
\affil[11]{ \orgname{Great Ormond Street Hospital for Children}, \orgaddress{\city{London}, \country{United Kingdom}}}

\abstract{Cellular immunotherapies are one of the mainstream cancer treatments unveiling the power of the patient's immune system to fight tumors. CAR T-cell therapy, based on genetically engineered T cells, has demonstrated significant potential in treating hematological malignancies, including B-cell lymphomas. This treatment has complex longitudinal dynamics due to the interplay of different T-cell phenotypes (e.g. effector and memory), the expansion of the drug and the cytotoxic effect on both normal and cancerous B-cells, the exhaustion of the immune cells, the tumor immunosupressive environments, and more. Thus, the outcome of the therapy is not yet well understood leading to a variety of responses ranging from sustained complete responses, different types of partial responses, or no response at all. We developed a mechanistic model for the interaction between CAR T- and cancerous B-cells, accounting for the role of the tumor morphology and metabolic status. The simulations showed that lesions with irregular shapes and high proliferation could contribute to long term progression by potentially increasing their immunosuppressive capabilities impairing CAR T-cell efficacy. We analyzed 18F-FDG PET/CT imaging data from 63 relapsed/refractory diffuse large B-cell lymphoma receiving CAR T-cells, quantifying radiomic features including tumor sphericity and lesion aggressiveness through standardized uptake values (SUV). Statistical analyses revealed significant correlations between these metrics and progression-free survival (PFS), emphasizing that individual lesions with complex morphology and elevated metabolism play a critical role in shaping long-term treatment outcomes. We demonstrated the potential of using data-driven mathematical models in finding molecular-imaging based biomarkers to identify lymphoma patients treated with CAR T-cell therapy having higher risk of disease progression.}


\maketitle

\section{Introduction}\label{sec1}
B-cell lymphomas are a diverse group of malignancies originating from B lymphocytes, characterized by significant variability in molecular profiles, clinical progression, and therapeutic responses. This includes both aggressive subtypes, such as diffuse large B-cell lymphoma (DLBCL), and more indolent forms like follicular lymphoma, each requiring tailored treatment strategies \cite{zelenetzNCCNGuidelinesInsights2023}. Certain lymphoma subtypes may occasionally present with features that resemble solid tumors. DLBCL, the most common and aggressive form, often starts in lymph nodes, but may involve extranodal sites such as the gastrointestinal tract, brain, or skin. Other subtypes, like primary mediastinal B-cell lymphoma and marginal zone lymphomas, can manifest as solid tumors in specific organs, further complicating treatment strategies \cite{chenCurrentChallengesTherapeutic2024}.

Traditional therapies for B-cell lymphomas, including chemotherapy, monoclonal antibodies (e.g., rituximab), and stem cell transplantation, offer limited options for patients with relapsed or refractory (r/r) disease. Frontline treatment for newly diagnosed aggressive NHL, such as R-CHOP (rituximab, cyclophosphamide, doxorubicin, vincristine, and prednisone), results in a cure for approximately 66\% of cases. However, outcomes for r/r aggressive B-cell lymphomas remain poor with salvage chemotherapy, which shows an overall response rate (ORR) of only 20\%-30\% and a median overall survival (OS) of approximately six months \cite{denlingerCARTcellTherapy2022}.

Cellular immunotherapies have revolutionized cancer treatment by harnessing cancer patient's immune system to specifically target and eliminate cancer cells. Chimeric antigen receptor T-cell (CAR T-cell) therapies have represented a significant step forward in the field. These treatments involve engineering T-cells obtained from patients to express receptors that recognize and bind to tumor-associated antigens on cancer cells. Upon \textit{in vitro} expansion and reinfusion into the patient, CAR T-cells recognize antigen-presenting cancer cells leading to their expansion and a cytotoxic action against the target cells  \cite{labaniehCARImmuneCells2023}. This therapy has shown remarkable success in treating hematological cancers such as leukemia and lymphoma \cite{savoldoCARCellsHematological2024} and the concept is being extended to solid non-hematological malignancies \cite{albeldaCARCellTherapy2024}.

The advent of anti-CD19 CAR T-cell therapy has revolutionized treatment for r/r B-cell lymphomas, offering significantly improved outcomes for patients who failed conventional therapies. FDA-approved CAR T-cell products, such as axi-cel (Yescarta), tisa-cel (Kymriah), and liso-cel (Breyanzi), have demonstrated impressive response rates in heavily pretreated patients. Across pivotal studies, ORRs for CAR T-cell therapy in r/r aggressive B-cell lymphomas range from 52\% to 83\%, with complete response (CR) rates of 40\% to 58\%. Many patients who achieve CR experience durable remissions, with a subset remaining progression-free for several years \cite{cappellLongtermOutcomesFollowing2023}.

Despite its success, CAR T-cell therapy faces several challenges, particularly when applied to solid-like B-cell lymphomas. The immunosuppressive tumor microenvironment (TME), which includes myeloid-derived suppressor cells (MDSCs), regulatory T-cells (T$_{\text{reg}}$), tumor-associated macrophages (TAMs), TGF-$\beta$, and immune checkpoint molecules, dampens immune responses and reduces CAR T-cell efficacy \cite{usluBloodExpandingCAR2024}. These obstacles are significant for the application of CAR T-cell therapy in solid tumors.

Mathematics provides advanced computational modeling techniques that facilitate the analysis of complex biological systems, allow simulating treatments in-silico, and predicting patient-specific responses to cancer \cite{altrockMathematicsCancerIntegrating2015}. These models may provide essential insights into key challenges in oncology, such as tumor dynamics and treatment resistance. By simulating tumor growth and the evolution of drug-resistant clones, it is possible to predict the long-term outcomes of various treatment strategies, leading to more effective and personalized care. In addition, several innovative applications of mathematics may reshape cancer treatments. The first is the development of computational simulations of in-silico trials that mimic clinical trials. This enables testing treatment strategies in digital cohorts before applying them to real patients, reducing cost, time, and ethical concerns associated with traditional trials \cite{insilico2024, PLOS2019}. The second concept, which is the so-called ``virtual twins'', are patient-specific mathematical models that integrate clinical, genomic, and imaging data to create a digital representation of an individual’s disease. Virtual twins aim to predict a patient’s response to specific treatments, allowing clinicians to personalize therapy plans and minimize adverse effects, ultimately improving patient outcomes \cite{digitaltwin2024}.

In this study, we built a compartmental mathematical model describing the interaction of CAR T-cells with target cancer cells in a solid tumor mass. Our model accounts for the biological aggressiveness, the expansion of the CAR T-cells and the cancerous cells, as well as the interacting mechanisms between both populations. The model was parametrized using a retrospective cohort of r/r DLBCL patients treated with CAR T-cells. Model simulations showed, and real-world data confirmed, the importance of the tumor aggressiveness combined with the lesion morphological complexity in determining the outcome of the therapy. 

\section{Results}\label{sec:results}

\subsection{Model simulations}\label{sec:results_math}

\begin{figure}[H]
\centering
  \includegraphics[width=0.8\linewidth]{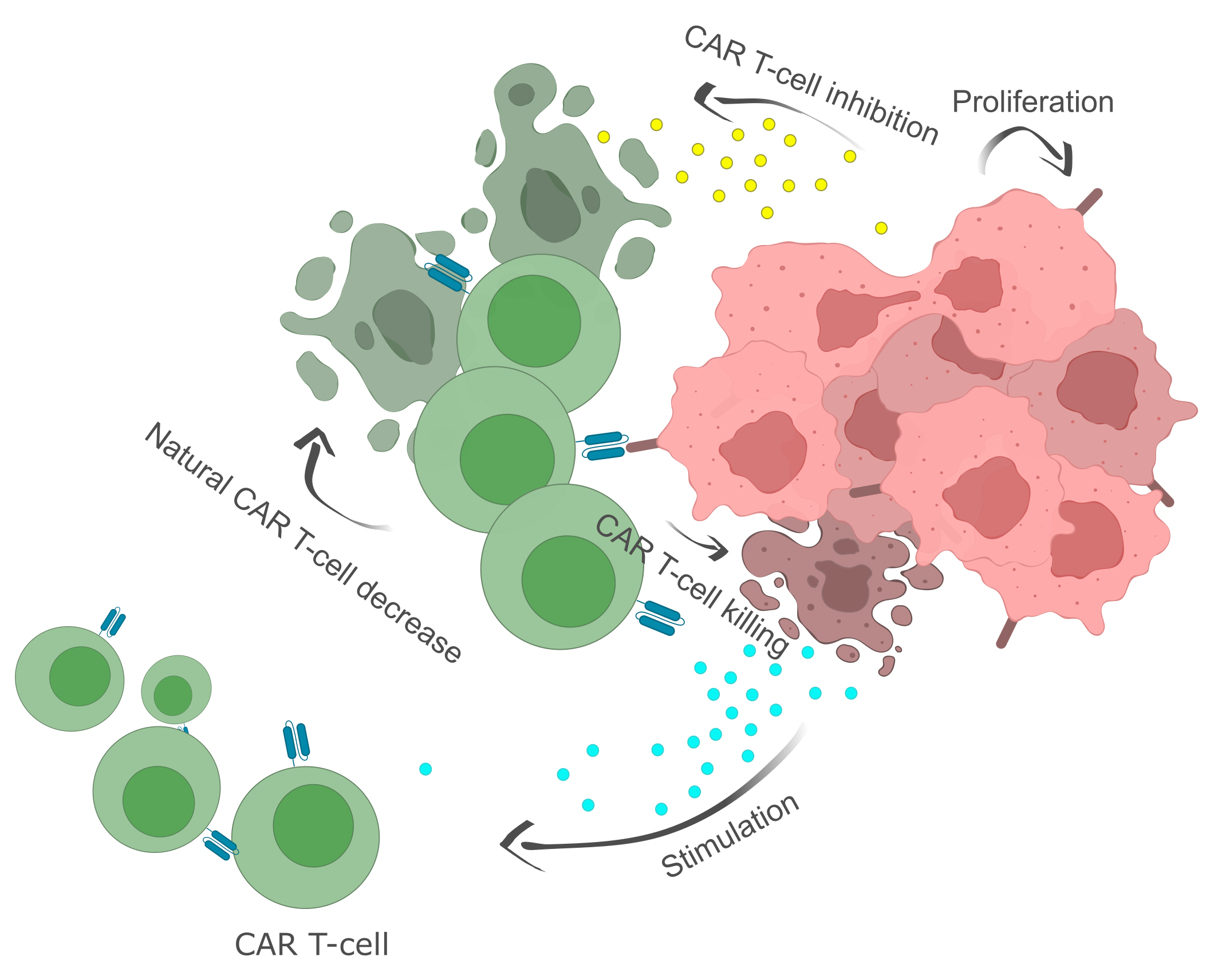}
  \caption{Schematic representation of the dynamics governing the interaction between CAR T-cells (green) and tumor cells (red), expressed in Eqs. \eqref{eq:model_full_1}  (see Sec. \ref{sec:methods_math} for details). CAR T-cells engage and kill tumor cells and proliferate due to direct stimulation and to the secretion of pro-inflammatory cytokines. Those cells get inactivated after a given period of time and also due both to the action of the tumor immunosupressive microenvironment and their inactivation in engaments with tumor cells. All of these processes are accounted for in the compartmental mathematical model.}
  \label{fig:scheme}
\end{figure}

We conducted simulations of a lymphoma lession growth using a two-compartment ordinary differential equations (ODE) model that incorporated the interaction between CAR T-cells ($C$) and CD19$^+$ cancerous B-cells ($B$)  (Figure \ref{fig:scheme}). The model basic rules as schematized in Fig. \ref{fig:scheme}. Essentially it was based on the simplifying assumption that all cancer cells express the CD19 antigen and are thus, susceptible to being targeted by CAR T-cells. Interactions between engineered T-cells and cancerous cells can lead either to the elimination of cancer cells and stimulation of CAR T-cell growth or inhibition of CAR T-cells (we refer to the latter effect collectively as immunosuppression). Cancer cells were assumed to grow logistically in the absence of T-cells.  Activated modified T-cells were considered to have have a finite lifetime. Finally, the interactions between effector and target cells occur mainly on the tumor surface as inner parts of the tumor correspond to regions difficult to access where immunosuppressive mechanisms are stronger. A detailed description of the model, including the parameter estimation, is given in Sec. \ref{sec:methods_math}. 

\begin{figure}[H]
\centering
\begin{subfigure}[b]{0.9\linewidth}
     \includegraphics[width=\linewidth]{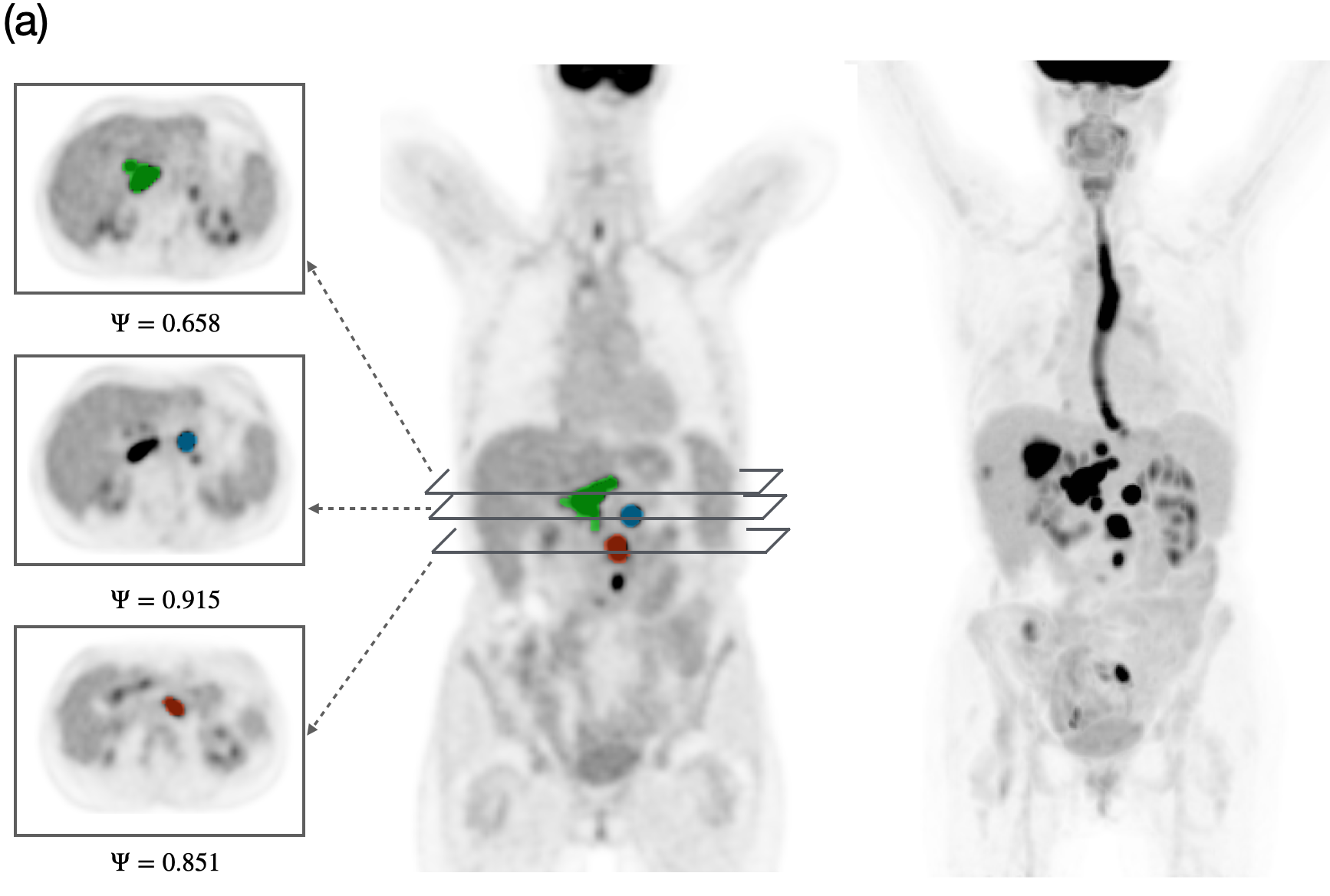}
\end{subfigure}
\begin{subfigure}[b]{0.9\linewidth}
    \includegraphics[width=\linewidth]{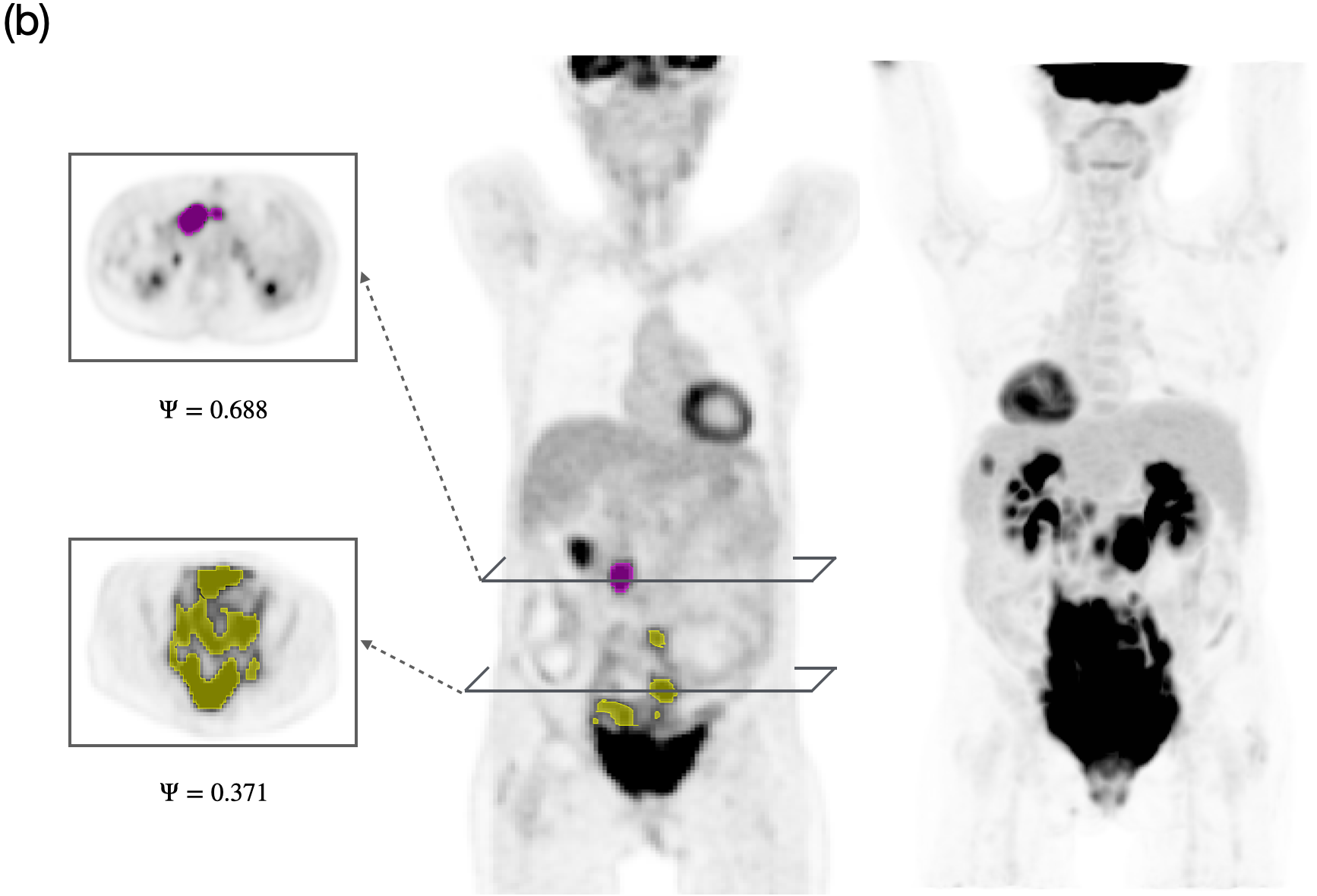}
\end{subfigure}   
  \caption{\textbf{PET/CT images of two different patients showing lesions with varying sphericity values ($\Psi$).} (a) 3D view of a full-body $^{18}$F-FDG PET scan of a lymphoma patient (rightmost image), with the coronal PET scan displaying three lesions of different sphericities, as shown in axial PET slices (leftmost images). (b) 3D view of a full-body PET scan of another lymphoma patient, with the coronal PET scan showing two lesions with different sphericities, along with several  axial slices. Higher $\Psi$ values indicate more spherical lesions, while lower values correspond to irregularly shaped lesions.} 
  \label{fig:lesion_spher}
\end{figure}

Details of model parameters, as well as their units and values used throughout this study are given in Table \ref{tab:model_full}. 
We computationally studied the long-term outcome of the treatment by varying the tumor proliferation rate $a$ and tumor lesion sphericity $\Psi$ within ranges estimated from literature and our own dataset. 
Figure \ref{fig:lesion_spher} shows representative examples of sphericity values observed in $^{18}$F-FDG PET images of different lesions from lymphoma patients included in the study. This visualization highlights the morphological heterogeneity among lesions, suggesting that lesion geometry could be a potential factor influencing treatment response and long-term therapy outcome.

For our simulations, we considered an initial tumor burden of $B_0 = 6.598 \times 10^9$ cells and set the initial number of engineered T-cells at the lesion site to $C_0 = 2.020 \times 10^8$ cells. These are the median estimated values from our dataset  (see Sec. \ref{sec:methods_par_estimate} for details).

\begin{figure}[H]
\centering
  \includegraphics[width=\linewidth]{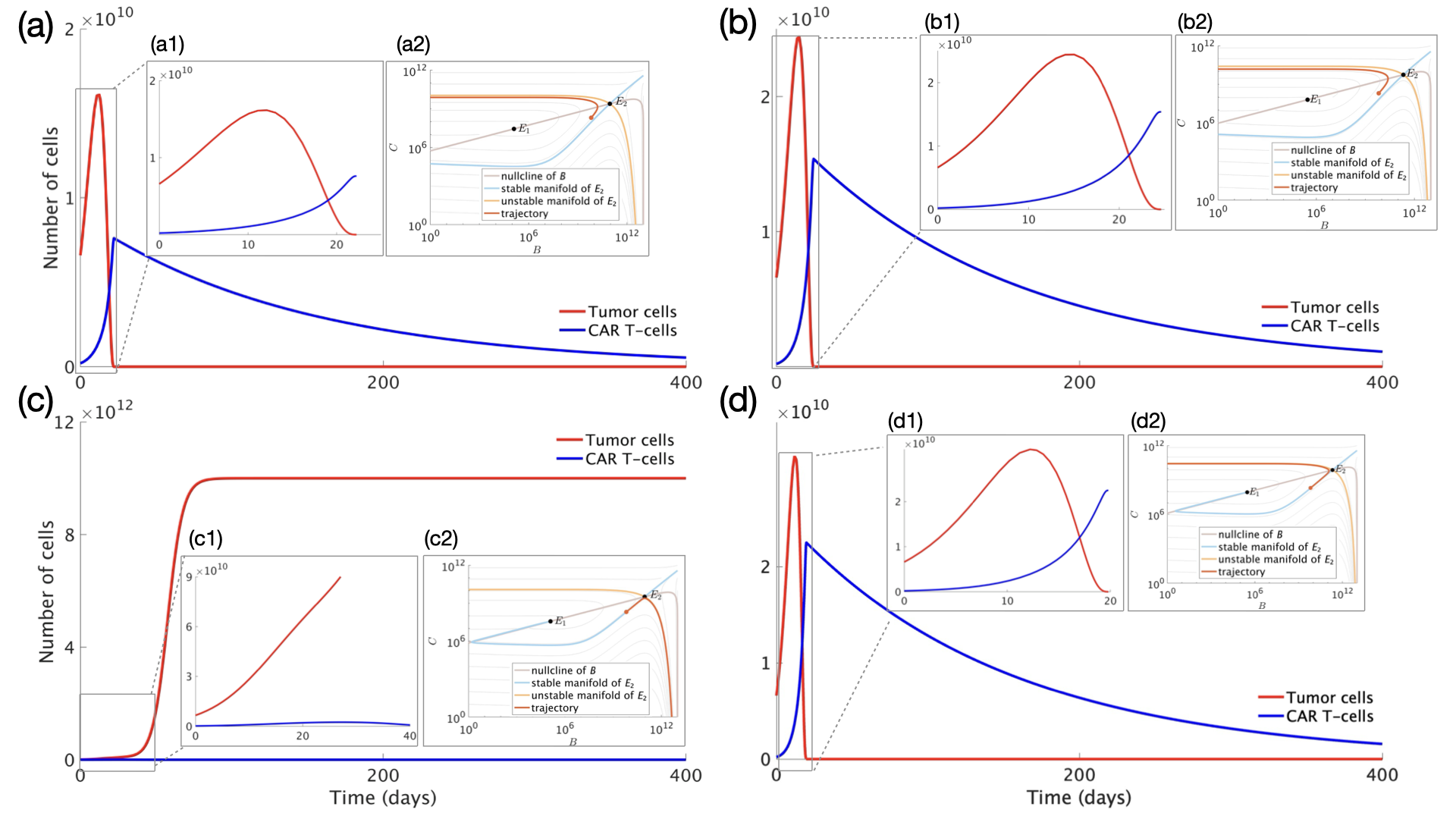}
  \caption{\textbf{Simulated tumor and CAR T-cell dynamics for different proliferation rates and sphericity values.} 
    (a) Long-term evolution over 400 days and (a1) short-term evolution over 25 days of tumor cells (red) and CAR T-cells (blue) for a proliferation rate of $a = 0.15$ day$^{-1}$ and lesion sphericity $\Psi = 0.4$. 
    (a2) Corresponding phase portrait of the system given by Equation \eqref{eq:model_full_1a} and \eqref{eq:model_full_1b}  emphasizing the reduction of the tumor cell population.
    Figures in (b), (c), and (d) describe the parameter combinations $(a = 0.15$ day$^{-1}, \Psi = 0.7),  (a = 0.2$ day$^{-1}, \Psi = 0.4)$, and $(a = 0.2$ day$^{-1}, \Psi = 0.7)$, respectively as in panel (a). The corresponding short-term dynamics as well as the phase portraits are given in (b1) and (b2), (c1) and (c2), and (d1) and (d2), respectively.}
  \label{fig:phaseplane_all}
\end{figure}

CAR T-cell treatment primarily relies on the interaction between the drug and antigen-presenting tumor cells. Intuitively, one might expect that a lower sphericity of the tumor lesion would result in a better disease outcome, as less spherical lesions might provide greater surface area and accessibility for CAR T-cells to engage with tumor cells. However, as illustrated in Fig. \ref{fig:phaseplane_all}, two lesions with identical initial tumor burden and proliferation rate, but differing sphericity values, exhibit distinct progression dynamics in silico. For instance, by comparing Figs. \ref{fig:phaseplane_all}(c) and (d), we observe that the lesion with lower sphericity progresses to an uncontrolled state reaching the tumor carrying capacity (i.e. $1/b$) at a higher proliferation rate \(a = 0.2\) day$^{-1}$. However, the lesion with a larger sphericity exhibits more balanced tumor-immune interactions, where \(B(t)\) is reduced to a clinically undetectable as examplified in Fig. \ref{fig:phaseplane_all} (d).

This behavior can be explained by the initial disease state’s location in the phase plane and the number of initial CAR T-cells successfully trafficked to the tumor site. As shown in the phase plane representations of the mathematical model given by Eqs. \eqref{eq:model_full_1} in Fig. \ref{fig:phaseplane_all}, an increase in proliferation rate shifts the initial disease state of the lower-sphericity lesion (Fig. \ref{fig:phaseplane_all}(c2)) into a region that trends toward an uncontrolled state. Conversely, this shift does not occur for the lesion with higher sphericity (Fig. \ref{fig:phaseplane_all}(d2)), which remains within a region conducive to controlled tumor-immune interactions. 

Model simulations demonstrate that for the given set of initial values, the outcome of the simulation is sensitive to both the proliferation rate $a$ and the lesion sphericity $\Psi$. We observe that the initial value is located near the lower branch of the unstable manifold of the saddle equilibrium $E_2$, which separates the basin of attraction for equilibrium $E_3 = (\bar{B}_3, \bar{C}_3) = (b^{-1}, 0)$ (escape scenario) from the rest of the system. Here, \( b \) denotes the inverse of the carrying capacity of the tumor cell population, such that \( \bar{B}_3 = b^{-1} \) corresponds to the maximal sustainable tumor burden in the absence of CAR T-cell control. The steady state $E_3$ of the system is characterized by a dominance of the malignant B-cells and a complete disappearance of the CAR T-cells. This separation is represented by the region on the right-hand side of the phase plane.
By increasing the proliferation rate (Fig.  \ref{fig:phaseplane_all}),  we observe that the lower branch of the unstable manifold of $E_2$ ``shifts'' upwards. For a lesion with a larger sphericity (0.7) (Fig. \ref{fig:phaseplane_all} (b2) and (d2)), the location of the initial value remains on the left-hand side of the lower branch of the unstable manifold of $E_2$, ensuring that the number of B-cells eventually decreases to a clinically undetectable state. However, for a lesion with the same initial state but a smaller sphericity ($\Psi = 0.4$) (Figs. \ref{fig:phaseplane_all}(a2,c2)), increasing the proliferation rate from 0.15 to 0.2 day$^{-1}$ results in a shift in the qualitative location of the initial state. In the latter case, the trajectory escapes and the tumor grows without being controlled by the drug.

\begin{figure}[h!]
  \centering
    \includegraphics[width=\linewidth]{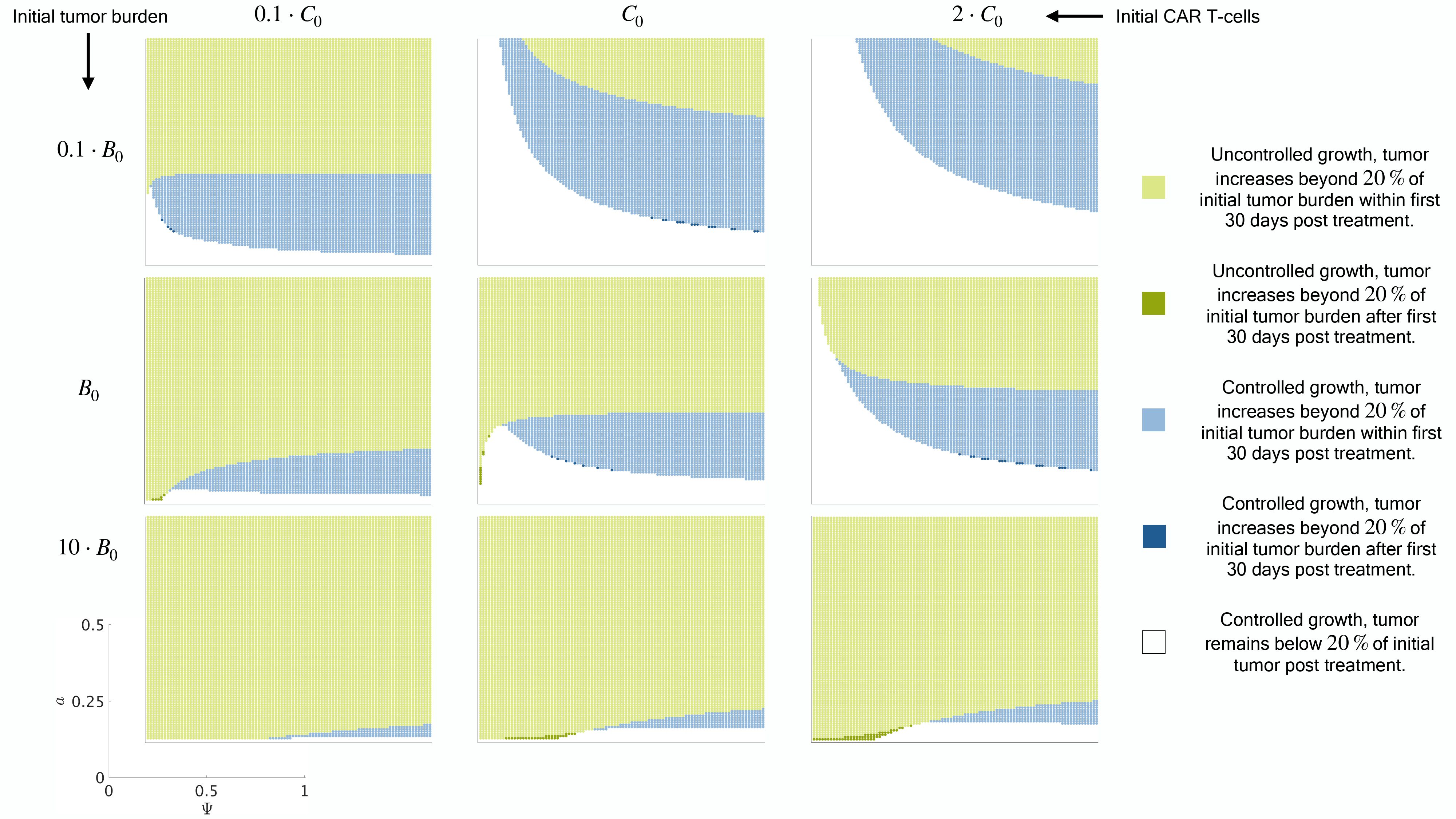}
  \caption{\textbf{Impact of proliferation rate ($a$), lesion sphericity ($\Psi$), initial CAR T-cell dose, and initial tumor burden on the dynamics of tumor cell population ($B$).} Each row represents a different initial tumor burden (\(0.1B_0\), \(B_0\), \(10B_0\), respectively from top to bottom), while each column shows the disease outcome for a different initial CAR T-cell dose (\(0.1C_0\), \(C_0\), \(2C_0\), respectively from left to right). $B_0$ and $C_0$ are obtained from the population median as given in Table \ref{tab:model_full} as well as all other parameter values. Light green dots represent parameter combinations of $a$ and $\Psi$ for which $B$ converges asymptotically to the carrying capacity, with tumor growth exceeding 20\% of its initial size within the first 30 days post-CAR T-cell infusion. Dark green dots indicate cases where this threshold is exceeded only after 30 days. Light blue dots correspond to parameter combinations of $a$ and $\Psi$, where $B$ declines to below 10 cells within 5000 days, but temporarily surpasses 20\% of its initial size within the first 30 days. Dark blue dots indicate cases, where the 20\% of the initial tumor burden is only exceeded after 30 days. Note that the time period considered for simulations exceeds the 3 years follow-up period in this study. Uncolored regions represent parameter combinations, where the tumor cell population reduces to a clinically undetectable state (below 10 cells in this case) without exceeding 20\% of the initial tumor burden. Note that only a few sets of parameter combinations lead to an increase of the tumor cell population above 20\% of its initial burden only after 30 days post CAR T-cell infusion.} 
  \label{fig:aPsi_overall}
\end{figure}

Figure \ref{fig:aPsi_overall} illustrates the parameter combinations of the proliferation rate $a$ and lesion sphericity $\Psi$ that lead to uncontrolled tumor growth (green), where the tumor cell population $B$ converges to its carrying capacity over a time course of 5000 days indicating treatment failure, for different initial tumor burden and CAR T-cell doses.

Parameter combinations highlighted in blue represent cases, where the tumor cell population $B$ eventually reduces to a clinically undetectable state, but temporarily exceed 20\% of its initial size during the time course of simulation. The 20\% threshold was chosen as a quantitative criterion to define transient tumor progression, capturing a meaningful temporary increase in tumor burden that may signal a loss of early therapeutic control \cite{kimmelRolesCellCompetition2021}. This definition offers a mathematically tractable and reproducible framework for modeling, in contrast to clinical definitions of progression, which are typically less precise and rely on qualitative evaluations of disease worsening relative to the nadir. Light blue and light green colors indicate the parameter pairs for which growth beyond 20\% of the initial tumor burden occurs within the first 30 days post CAR T-cell infusion.
Using the median estimates for $C_0$ and $B_0$ across the 63 patients, we observed that the range of values for the parameter $a$ corresponding to an uncontrolled disease outcome (colored in green), increases with decreasing sphericity $\Psi$ as shown in the middle plot. 
However, as we double the initial number of CAR T-cells (i.e. to $4.040 \times 10^8$ cells) while keeping $B_0$, we observed a change in the shape of the colored regions indicating a better treatment outcome even for aggressive nonspherical tumors (low $\Psi$ and high $a$). This trend was also observed when computationally lowering the initial tumor burden (i.e. to $0.1 B_0$) and high initial CAR T-cell doses (i.e. $C_0$ and $2 C_0$). 

These findings based on computational simulations emphasize the importance of CAR T-cell dose, lesion geometry and tumor proliferative potential as variables potentially influencing the outcome of the dynamics, with more initial CAR T-cell successfully trafficked to the tumor lesion site potentially lead to a better treatment outcome. Moreover, model simulations also showed that for nonspherical tumor lesions, the outcome after treatment would not depend on the threshold used to define PD. However, for a more regular lesion, the lower the PD threshold is, the larger the range of tumor cell proliferation rate $a$ is for which the disease becomes progressive (Fig. S1).
Furthermore, simulations show that in most of the cases, if the tumor cell population were to grow beyond 20\% of its initial size, this will most likely occur within the first 30 days post CAR T-cell infusion highlighting the potential importance of frequent monitoring of the disease within the first 30 days post-treatment.

A detailed mathematical analysis of the existence and stability of equilibrium states are provided in Supplementary Information (SI), which offers deeper insights into the long-term behavior of the system and provides a theoretical foundation for interpreting the results of the simulations.

\subsection{Imaging features ruling response}\label{sec:results_stat}

Our mathematical model suggests that lesions with high proliferation rates and complex morphologies combine both biological aggressiveness and immunosuppressive capabilities, leading to an undesirable outcome after the initial response to therapy, thereby reducing PFS.

To evaluate the model predictions, we analyzed the cohort of 63 patients, used to parametrize the model and described in Sec. \ref{sec:methods_patient}, diagnosed with aggressive B-cell non-Hodgkin lymphoma, and treated with commercial CAR T-cell therapies. Baseline $^{18}$F-FDG PET scans were available for all patients prior to CAR T-cell infusion. Tumor lesions were segmented to quantify the initial tumor burden and inform model parameters. Additional details regarding cohort, imaging acquisition, and processing are provided in Sec. \ref{sec:methods_patient}. Morphological descriptors, including sphericity, were obtained for each individual lesion and proliferation levels obtained from metabolic activity metrics (SUV), since glucose uptake is mostly used for biosynthesis and is thus a surrogate of proliferation. For each patient, PD was assessed by clinicians using the Lugano Classification, which evaluates metabolic response based on PET/CT imaging and the Deauville Score \cite{TAMAYO2017312}. In the context of CAR T-cell therapy, additional clinical assessments, including biomarkers such as lactate dehydrogenase (LDH) and patient symptoms, were also obtained and considered to identify progressions.

\begin{figure}[H]
    \includegraphics[width=\linewidth]{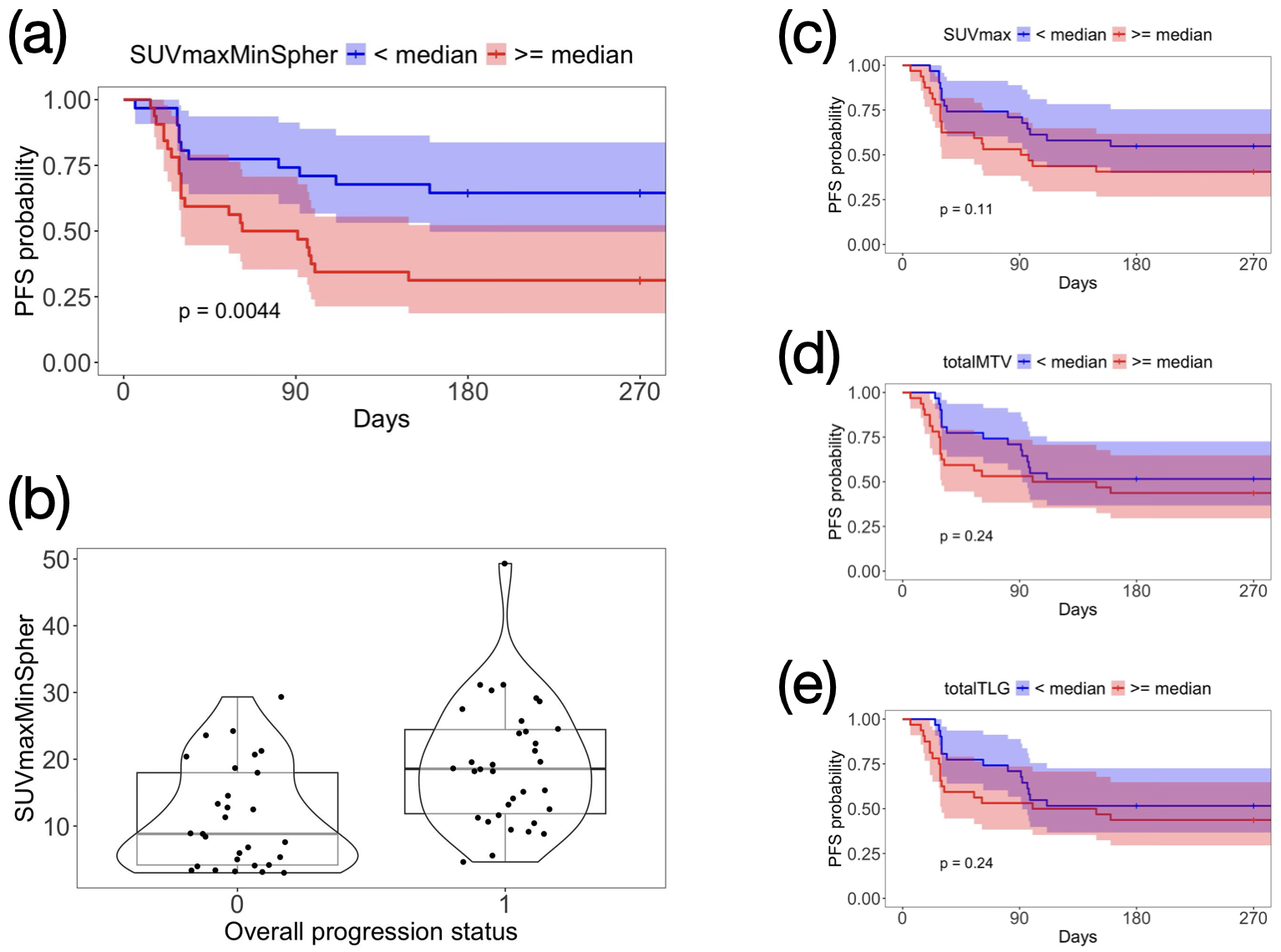}
  \caption{\textbf{Statistical analysis of the most relevant metabolic features studied.} Panels (a), (c), (d), and (e) show Kaplan-Meier survival curves based on median stratification for SUVmaxMinSpher (median = 14.131), SUVmax (median = 18.630), total metabolic tumor volume (MTV, median = 127.122cm\(^3\)), and total tumor lesion glycolysis (tTLG, median = 804.892 g), respectively. Colored regions represent the 95\% confidence intervals.
  Panel (b) shows the distribution of SUVmaxMinSpher according to patients' overall progression status, where 1 indicates disease progression during follow-up.}
\label{fig:SUVmaxMinSpherBoxPlot}
\end{figure}

To study the relation between these variables and PFS we developed a multivariate model incorporating the key metabolic features: metabolic tumor volume (MTV), total lesion glycolysis (TLG), and maximum standardized uptake value (SUVmax), as well as morphological features of lesions with both the smallest and largest sphericity, along with their corresponding volumes and SUVmax values. The detailed procedure can be found in SI. The most relevant variable in the univariate analyses was
 the SUVmax of the lesion with the lowest sphericity (SUVmaxMinSpher). Patients with larger values of that measure experienced significantly poorer PFS, with a particularly steep decline observed within the first 45 days post infusion. 
 Fig. \ref{fig:SUVmaxMinSpherBoxPlot}(a) shows the Kaplan-Meir plots (p=0.0044) using the median to separate the groups.  
 Figure \ref{fig:SUVmaxMinSpherBoxPlot}(b) illustrates the distribution of SUVmaxMinSpher values in patients with and without overall progression. A Wilcoxon rank-sum test yielded a p-value of 0.0003, indicating a statistically significant difference (on a level of 0.05) between the two groups.
The complete analysis is provided in SI. Interestingly, as illustrated in Fig. \ref{fig:SUVmaxMinSpherBoxPlot}(c,d,e), none of the classical metabolic parameters totalMTV, totalTLG, SUVmax showed a statistically significant association with PFS (p-values of 0.24 , 0.24  and 0.11 respectively).

\section{Discussion}\label{sec:discussion}

The standardization of high-resolution medical imaging for tumor diagnosis, treatment planning, and follow-up has opened up the possibility to use imaging-derived quantitative metrics as biomarkers of disease prognosis and treatment response. Specifically tumor morphology, as assessed by high-resolution medical imaging, has been shown to correlate with the disease outcome in different histologies. Shape related measures, such as tumor surface regularity, quantify the complexity of the tumor-normal tissue interface, and are thus related to the phenotypes creating the tumor invasion front. 

Higher sphericity of tumor lesions has been associated with good prognosis in glioblastoma, both in MRI \cite{perez-betetaTumorSurfaceRegularity2018, linTumourSurfaceRegularity2025} 
and $^{18}$F-fluorocholine PET/CT images \cite{garciavicente18FFluorocholinePETCT2019}, 
oral cavity cancer \cite{lucchiPretreatmentTumorVolume2023}, 
oral cavity squamous cell carcinoma \cite{tarsitanoPretreatmentTumorVolume2019},
invasive breast cancer patients receiving neoadjuvant chemotherapy \cite{liTumorSphericityPredicts2020},
brain metastases receiving interstitial thermal therapy \cite{sanvitoSmallPretreatmentLesion2023},
pharynx squamous cell carcinomas \cite{fujimaIntegratingQuantitativeMorphological2018},
locally advanced head and neck carcinomas treated with neoadjuvant immunotherapy \cite{vanderhulstQuantitativeDiffusionWeightedImaging2022},
hepatocellular carcinomas \cite{wangThreedimensionalMorphologyScoring2024},
ovarian cancer \cite{zuoPredictionOvarianCancer2024}, and
locally advanced cervical cancer treated with chemoradiotherapy \cite{pedrazaValueMetabolicParameters2022}.
Sphericity was noticeably higher for low-risk thymomas compared to high-risk thymomas \cite{yamazakiQuantitative3DShape2020},
and also higher for lower grade pancreatic neuroendocrine tumors than for their higher grade counterparts \cite{choiPancreaticNeuroendocrineTumor2018}.
High sphericity predicted good response to neoadjuvant immunotherapy in patients with locally advanced head and neck carcinoma \cite{vanderhulstQuantitativeDiffusionWeightedImaging2022},
and is associated with delayed progression in nonmetastatic nasopharyngeal carcinoma after intensity-modulated radiation therapy \cite{duRadiomicsModelPredict2019}.
Complex surface/contour patterns and/or low sphericity have also been associated with worse disease outcomes for patients with renal cell carcinoma \cite{zhuTumorContourIrregularity2024, xuPredictiveValueRenal2022, daiTumorContourIrregularity2021}.

Surface irregularity obtained from MRI images has been found to correlate with different metrics of invasiveness, such as microvascular invasion in hepatocellular carcinoma \cite{liPreoperativeThreeDimensionalMorphological2024},
extranodal extension status in tongue squamous cell carcinoma \cite{yangDeterminationCervicalLymph2021},
presence of liver metastases in castration-resistant prostate cancer patients \cite{mattoniPSMAPETEvaluation2022},
and biological aggressiveness in pituitary adenomas \cite{wangShapeTextureAnalyses2023}.
Most likely due to this more invasive behavior of irregular cancers, irregularity has been reported to be a predictor of recurrence after surgery in 
glioblastoma \cite{perez-betetaTumorSurfaceRegularity2018}
and low-risk renal cell carcinoma \cite{xuPrognosticValueTumour2023}.
 
In the context of DLBCL some works have related morphological and volumetric measures with outcome. Sphericity was found to be significant in a multivariate model derived from a 33-patient cohort receiving first-line chemoimmunotherapy \cite{ferrer-loresPrognosticValueGenetic2023}. \citet{sheng18FFDGPETCT2024} analyzed a cohort of 90 DLBCL patients treated with CAR T-cell therapy and found that tumor size predicted treatment success. Patients with smaller tumors (maximum diameter $<6$ cm) had higher complete response rates, improved PFS, and 
better overall survival compared to those with larger tumors (maximum diameter $\geq 6$ cm). In particular, larger tumors were associated with more immunosuppressive TME, marked by increased M2 macrophages, cancer-associated fibroblasts, and T-cell exhaustion, all of which impaired the efficacy of CAR T-cell therapy. However, to the best of our knowledge, the role of tumor morphology has not been studied in the context of the response to CAR T-cell treatments. 

Mathematical models have been widely used to study the dynamic interactions between CAR T-cells and tumor cells in both hematological and solid tumors. However, few models incorporate tumor geometry, despite its significant impact on treatment response. In B-cell lymphomas, the geometry of the tumor is especially important due to the clustered and often irregular arrangement of malignant lymphoid cells within lymph nodes or extranodal tissues. This structure can potentially impede immune cells' access and reduce therapeutic effectiveness. 

In this study, we developed a novel compartmental mathematical mechanistic model that incorporated the two key interacting populations: CAR T-cells and lymphoma B-cells. Building upon the foundation of Kuznetsov's model \cite{kuznetsovNonlinearDynamicsImmunogenic1994}, a well-established framework for studying tumor-immune interactions, we extended the model to integrate tumor sphericity as a critical parameter. While the original model primarily focused on the kinetics of tumor-immune interactions, our approach captured in the most simple way how tumor shape, particularly deviations from sphericity, can influence CAR T-cell infiltration, immune supression, spatial distribution, and cytotoxic activity. 
Our framework was designed to computationally study the interplay of cancer aggressiveness, measured in terms of proliferation rate, with the morphological aspects of the interaction. We aimed to enhance our understanding of the factors that influence the outcomes of CAR T-cell therapy and study the potential relevance of the tumor spatial geometry in treatment strategies, particularly for B-cell lymphomas.

Using the patient cohort included in this study to parametrize the model, we observed that tumor lesions with both low sphericity and high tumor proliferation rate would exhibit poor responses to CAR T-cell therapy, compared to lesions of the same size and proliferation rate with higher sphericity. 
Even though lesions with low sphericity, corresponding to highly irregular or asymmetrical shapes, are more exposed to the interaction with CAR T-cells, the lesion irregularity poses a further challenge for CAR T-cells to infiltrate and navigate within the spatially complex structured TME.
This uneven penetration may directly result in insufficient engagement with tumor cells, leaving significant portions of the lesion untreated. Combined with a high tumor proliferation rate, indicative of aggressive tumor behavior, the lesion may establish a stronger immunosuppressive environment, further limiting the efficacy of the treatment.
Simultaneously, a high tumor growth rate can outpace the cytotoxic activity of CAR T-cells.

In this study, we showed the potential of integrating clinical data with mathematical modeling to find candidate biomarkers that may be used in real-world settings. It is remarkable that the model suggested to focus on individual lessions having low sphericity and high SUVmax, instead of considering global metrics of the disease. Interestingly, the scenario described in the model was fully confirmed by the results obtained from clinical data, where Kaplan-Meir analysis for the progression-free survival highlighted the key role of the patient's highly metabolic and least spherical lesions.

Our study focused on DLBCL, the most common and aggressive form of lymphoma. It would be also interesting to study the role of the metrics derived for other lymphomas or even tumors of others histologies for which many trials involving CAR T-cells are in the works. The biomarkers found here could provide a guide in selecting patients for treatment and the mechanistic model provide hints on how to personalize treatment or improve the outcome.

A limitation of our study was the small size of the patient cohort included, and the fact that it was obtained from a single institution. A confirmatory study in a different cohort would be necessary to validate the observation made in our study.

Moreover, the patients' PET images were segmented using one single method. The robustness of the results should be tested using other routinely used segmentation methods including manual expert segmentation methods.
Furthermore, our mathematical model might need to be further modified in order to provide more personalized and accurate outcome prediction.
For instance, in modeling sphericity, we simplified the scenario by using a constant parameter $\Psi$ obtained from pre-CAR T-cell infusion imaging data. However, tumor growth is highly dynamic, resulting in a non-constant lesion sphericity over time. One possible approach would be to consider sphericity as a function of the number of tumor cells, assuming a larger tumor lesion to be associated with a more nonspherical shape. 
In addition, our model incorporates a parameter $q$ that represents the inhibition of CAR T-cells caused by tumor cells, reflecting the immunosuppressive effects and exhaustion of CAR T-cells. However, exhaustion of CAR T-cells does not necessarily lead to a complete shutdown of the cells' killing functionality. Instead, the corresponding term describes a functional but yet hypo-responsive state of cytotoxic T-cells \cite{crespoCellAnergyExhaustion2013, jiangExhaustedCD8+TCells2021}.
Finally, spatial aspects might play an important role and applying cellular automata models or mesposcopic simulation methods as used in \cite{bordel-vozmedianoGeometricImmunosuppressionCART2025, jimenez-sanchezMesoscopicSimulatorUncover2021} may provide additional insights.

\section{Methods}\label{sec:methods}

\subsection{Patients}\label{sec:methods_patient}

This retrospective study was approved on November 3rd, 2021, by the Institutional Review Committee (IRC) at Hospital Universitario La Paz under protocol number GETH-MATCART-2021.

A cohort of 63 patients with aggressive non-Hodgkin B-cell lymphoma was used, as detailed in Table \ref{tab:demographics}. The study adhered to a comprehensive protocol that encompassed patient selection criteria, drug information, data collection methods, and ethical considerations. Demographic, follow-up, and treatment response data were collected using REDCap, a secure web-based platform for clinical research run by the Spanish Group of Hematopoietic Transplant and Cellular Therapies (GETH). Patients were treated with CAR T-cell therapy and followed at the same institution. Among these patients, 73\% were diagnosed with diffuse large B-cell lymphoma (DLBCL) and others with high-grade or transformed B-cell lymphomas. Most patients (69\%) received Axicabtagene ciloleucel (axi-cel), while 31\% were treated with Tisagenlecleucel (tisa-cel), both CAR T-cell therapies are FDA- and EMA-approved.

In addition to clinical assessments, CAR T-cell quantification was performed by hematologists using polymerase chain reaction (PCR) and flow cytometry (CF) at different time points throughout the treatment. These data were also recorded in REDCap, contributing to the comprehensive tracking of patient responses and treatment progress.

Treatment responses were assessed at intervals of 1, 3, 6, 9, 12, 18 months, and at 2 and 3 years post-infusion. Responses recorded at each time point were considered valid until the next evaluation. Disease progression was documented with exact dates, and PFS was defined as the time from CAR T infusion to relapse or progression. 

Patients lost to follow-up due to relocation or logistical issues were handled by treating the last available evaluation as valid until the next scheduled assessment, thus becoming censored events in the analysis. This method ensured consistency in analyzing incomplete data.

\begin{table}[h!]
    \centering
    \caption{Patient baseline characteristics and response to treatment.}
    \renewcommand{\arraystretch}{1.5} 
    \begin{tabular}{|>{\raggedright\arraybackslash}m{4cm}|>{\raggedleft\arraybackslash}m{2cm}|} 
        \hline
        \textbf{Number of patients} & $63$ \\ 
        \hline
        \textbf{Median age (range), years} & 61 (32-79) \\		
        \hline
        \textbf{Gender} &  	\\	
        \hfill Male & 38 (60.3 \%)   \\ 
        \hfill Female & 25 (39.7 \%)   \\ \hline
        \textbf{Lymphoma subtype} &   \\
        \hline Diffuse large B cell lymphoma & 46 (73\%)   \\
      High grade B cell lymphoma with MYC and BCL2 and/or BCL6 rearrangements & 5 (7.9 \%)  \\
        Primary mediastinal (thymic) large B-cell lymphoma & 3 (4.7 \%)  \\
        Transformed follicular lymphoma & 5 (7.9 \%)  \\
        Composite lymphoma & 3 (4.7 \%) \\         \hline
        \textbf{Autologous stem cell transplant} & 10 (15.9 \%)  \\ 		        \hline
        \textbf{Number of previous lines of treatments} &   \\ 
          \hfill  $= 2$ & 53 (84.1\%) \\ 
          \hfill  $> 2$ & 10 (15.9 \%) \\ 	        \hline	
        \textbf{Bridge therapy} &  \\ 		
           \hfill Yes & 47 (74.6 \%) \\ 
          \hfill  No & 16 (25.4 \%) \\ 		        \hline
        \textbf{CAR T treatment type} &  \\ 		
          \hfill  Tisagenlecleucel & 20 (31.7 \%)  \\ 
          \hfill  Axicabtagene ciloleucel & 43 (68.3 \%) \\        
          \hline		
    \end{tabular}
    \label{tab:demographics}
\end{table}

All patients had r/r disease with 84.1\% receiving two prior lines of therapy and 15.9\% receiving more than two. Prior to CAR T treatment, 15.9\% of patients underwent autologous stem cell transplantation (ASCT) as a second line treatment.
Bridge therapy to manage disease during CAR T manufacturing was administered to 74.6\% of patients, predominantly with chemotherapy (61\%) or combinations of radiotherapy and immunotherapy. The remaining 25.4\% did not require bridge therapy due to disease stability or clinical contraindications.

Early response to CAR T therapy, assessed within three months, showed a 57.1\% response rate (36 patients), with 71\% achieving complete response (CR) within the first month. Among initial CR responders, 56\% sustained remission up to 18 months. Some partial responders (PR) converted to CR by day 90, highlighting the evolving nature of CAR T outcomes. However, 38.1\% experienced disease progression during early evaluation, with additional cases of late progression observed at six and 18 months post-infusion. Further details are provided in Figure S3 of the SI. Response evaluation within 90 days was not performed for one patient due to lack of PET images. 

\subsection{Image analysis}
\label{sec:methods_image}
\begin{figure}[h!]
\centering
  \includegraphics[width=\linewidth]{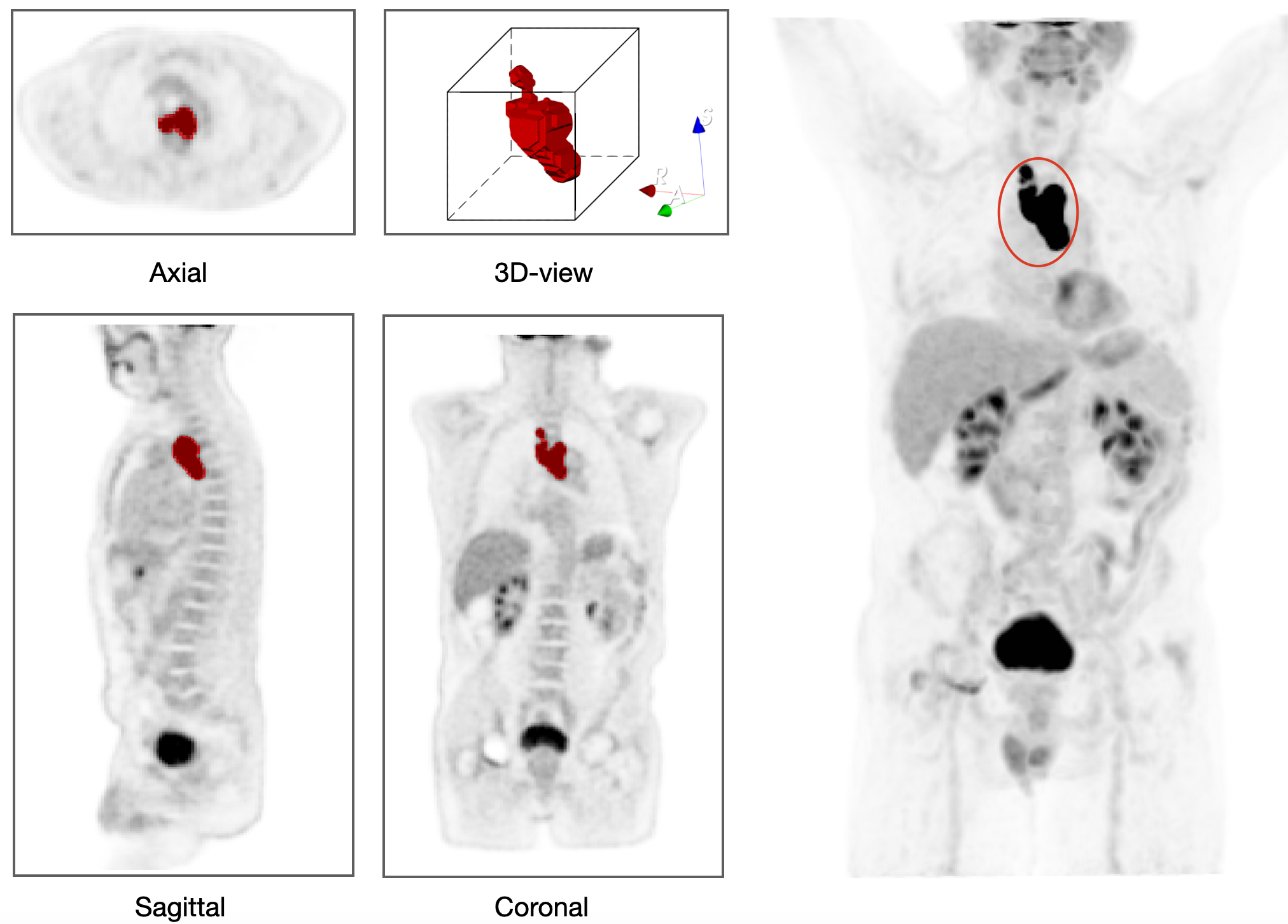}
  \caption{\textbf{Whole-body PET scan of a lymphoma patient and example of segmentation}. This patient had a single segmented lesion, which is determined using MIM software. Coronal, sagittal, and axial views of the lesion, along with a 3D rendering are shown in different panels. The segmented regions of interest (ROIs) delineate areas of increased metabolic activity, providing a comprehensive assessment of the lesion's morphology and spatial distribution.}
  \label{fig:pet}
\end{figure}

Pretreatment baseline $^{18}$F-FDG PET images were segmented by nuclear medicine doctors from the University
Hospital of Salamanca using MIM Encore software (Euro Automation S.L.). Segmentation criteria included a standardized uptake value (SUV) larger than 4 and a minimum lesion volume of 3 ml to ensure clinically relevant analysis. Figure \ref{fig:pet} illustrates an example of a baseline PET/CT image of a lymphoma patient with a single segmented lesion.

We extracted a range of quantitative imaging features from baseline PET images to assess tumor characteristics. We included all classical metabolic activity features, including total metabolic tumor volume (totalMTV), overall total lesion glycolysis (totalTLG), and maximum standardized uptake value (SUVmax).  Lesion shape features from the baseline PET scans were also computed using MATLAB (version 2024a) function \texttt{radiomics()} without any resampling, resegmenting, or additional discretizing of the input data. 
For each pixel of the segmented files, SUV values were computed from the DICOM data using the formula \cite{kinahanPositronEmissionTomographyComputed2010}:
\begin{equation*}
    \text{AC}_{\text{pix}} \cdot \frac{\text{W}}{\text{FDGdose}},
\end{equation*}
where $\text{AC}_{\text{pix}}$ is the activity concentration in a pixel in $\mathrm{Bq/ml}$, W is the body weight in g, and FDGdose is the decay corrected radiotracer administered in Bq.

TLG was derived by combining the MTV with the average standardized uptake value (SUVmean) of the tumor. We calculated totalTLG by summing up the TLG values across all lesions identified in the PET images. This approach allowed us to capture the extent of tumor volume while accounting for the metabolic activity within each lesion. 

In addition to these classical features, we expanded our analysis to include characteristics related to tumor shape, specifically focusing on sphericity given by
\begin{align}\label{eq:psi}
    \Psi = \dfrac{(36\pi)^{\frac{1}{3}} V^{2/3}}{A},
\end{align}
where $V$ describes the tumor volume and $A$ its surface area.
This non dimensional metric $\Psi$ compares the lesion surface area to the surface area of a sphere having the same volume. It ranges between 0 and 1, where a value of 1 indicates a sphere. The sphericity, which quantifies the surface structure of a tumor lesion, can provide insights into the biological behavior of the tumor.

For each patient, we analyzed all lesions individually to derive shape-related metrics. Specifically, we identified the lesion with the largest sphericity, termed maxSpher, and computed its volume (volMaxSpher) and maximum standardized uptake value (SUVmaxMaxSpher). Similarly, we identified the lesion with the smallest sphericity, referred to as minSpher, and assessed its volume (volMinSpher) and maximum standardized uptake value (SUVmaxMinSpher).

\subsubsection{Statistical methods}

Multivariate Cox regression were performed using the \texttt{survival} package in R. Features included those described in Section \ref{sec:methods_image}, i.e. totalMTV, totalTLG, SUVmax, maxSpher, volMaxSpher, SUVmaxMaxSpher, minSpher, volMinSpher, and SUVmaxMinSpher. Model selection was guided by the overall significance of the Cox regression model, the hazard ratios (HR) of individual covariates, p-values from the Wald test for each feature, and Spearman correlations to assess potential collinearity between covariates. To ensure model robustness and interpretability, some features were excluded (see SI). The final model was compared with an alternative model including the CAR T-cell product. 

Given the presence of competing risks -- specifically, patients who died without experiencing disease progression or relapse -- we performed a sensitivity analysis. In this analysis, we refitted the Cox model under a ``worst-case scenario'', assuming that patients who died without documented progression had, in fact, experienced progression at the time of their last follow-up. This approach helped mitigate potential bias introduced by competing risks and reinforced the robustness of our findings under different assumptions. Details of the procedure are provided in SI.

\subsection{Mathematical model}\label{sec:methods_math}

Inspired by the model used in \cite{kuznetsovNonlinearDynamicsImmunogenic1994}, we developed a mathematical model to describe the dynamic interactions between CAR T-cells ($C$) and lymphoma B-cells ($B$) of a single lesion over time.
We incorporated the sphericity of the tumor lesion, defined in \eqref{eq:psi}, where $\Psi = 1$ represents a perfect sphere, and lower values correspond to more irregular shapes. From this relationship, the tumor surface area can be expressed as being proportional to the product of the inverse of its sphericity 
and its volume raised to the power of two-thirds: $$ A \propto \spher \cdot V^{2/3}. $$  
Consequently, the inverse of the sphericity $1/\Psi$ lies within the range $[1, \infty)$, with larger values representing more irregular shapes.
In the context of B-cell lymphoma, the contact of tumor and immune system is limited to the tumor surface, which is proportional to $\spher \cdot V^{2/3}.$ We further assumed that the number of cancerous B-cells is proportional to the lesion volume and considered the case, where a positive amount of CAR T-cells have successfully trafficked to the tumor. The assumptions underlying our mathematical model are illustrated graphically in Figure \ref{fig:scheme}.

First, we assumed that all detected cancer cells are susceptible to elimination by CAR T-cells upon direct interaction. This is consistent with the mechanism of action of CAR T-cell therapy, where the engineered T-cells specifically target and destroy B-cells expressing the CD19 antigen.
Second, in the absence of CAR T-cells, the growth of the tumor cell population is governed by a logistic growth. This is a biologically plausible assumption that accounts for the limited availability of resources such as nutrients and space within the tumor microenvironment.
Next, interactions between cancerous B-cells and CAR T-cells were assumed to trigger signaling pathways, such as the release of cytokines, that result in the expansion of the T-cell population and recruitment of additional CAR T-cells to the tumor site. However, this recruitment is subject to an upper limit, beyond which CAR T-cell activation and recruitment saturate.
Fourth, in the absence of cancerous B-cells, CAR T-cells are assumed to undergo apoptosis, reflecting their natural life cycle in the absence of antigenic stimulation. This aligns with experimental observations showing the ability of self-regulation of T-cells in the absence of their target.
Fifth, each CAR T-cell has a finite capacity for interaction with cancerous B-cells. After a certain number of interactions, CAR T-cells may become inhibited. This is due to factors such as the limited production of effector proteins needed to kill tumor cells and the tumor's ability to induce apoptosis in nearby CAR T-cells.
Sixth, the interactions between CAR T-cells and tumor cells happens mainly at the tumor surface, that is the region most accessible to the immune cells. It is also known that T-cells have difficulties in accessing and extravasating into the deepest parts of the tumor that are also expected to have a stronger immunosupressive environment.
Finally, we assumed that other immune effector cells, such as natural killer (NK) cells or macrophages, have negligible contributions to the killing of cancerous B-cells in comparison to the T-cells, and are therefore, omitted in this model.

Putting these assumptions together and assuming Michaelis-Menten kinetics for the stimulated recruitment of CAR T-cells by tumor cells yields the system of differential equations

\begin{subequations}
\label{eq:model_full_1}
\begin{eqnarray}
    \frac{dB}{dt} & = & aB(1-bB) - \spher d CB^{2/3}, \label{eq:model_full_1a}\\
    \frac{dC}{dt} & = & -\spher q CB^{2/3} + \frac{\spher j B^{2/3}}{g ^{2/3} + \spher B^{2/3}}C - m C, \label{eq:model_full_1b}
\end{eqnarray}
\end{subequations}
where all parameters are assumed to be positive.
The stimulation term in the second equation can be written as
\begin{align*}
    \frac{\spher j B^{2/3}}{g ^{2/3} + \spher B^{2/3}}  = \frac{j B^{2/3}}{\left(\frac{g}{\Psi^{-3/2}} \right)^{2/3} + B^{2/3}} = \frac{j B^{2/3}}{\tildeg^{2/3} + B^{2/3}},
\end{align*}
where $\tildeg := g/\Psi^{-3/2} \leq g$, is the adjusted number of tumor cells at which the stimulated recruitment rate reaches its half maximum. Notice that $\tildeg$ decreases with the sphericity, which implies that for a more nonspherical lesion, the number of tumor cells that come into contact with CAR T-cells is increased leading to a smaller total amount of tumor cells for which the stimulated recruitment rate reaches its half maximum.

Equation (\ref{eq:model_full_1a}) describes tumor cell proliferation using a logistic growth term, where $a$ represents the intrinsic proliferation rate, and $b$ the inverse of the tumor carrying capacity. This logistic form reflects the biological constraint that tumor growth is limited by environmental resources or anatomical barriers. Tumor destruction by CAR T-cells is modeled using a lytic term, where $d$ quantifies the efficacy of CAR T-cell-mediated cytoxity.
Equation \eqref{eq:model_full_1b} models the dynamics of CAR T-cells, capturing their depletion through interactions with lymphoma B-cells, proliferation, and intrinsic decay. Two main biological processes behind depletion in the CAR T-cell population involve exhaustion and immunosuppression. Exhaustion occurs when the functionality of CAR T-cells is reduced after prolonged engagement with tumor antigens, resulting in diminished efficacy. Immunosuppression, on the other hand, arises from inhibitory signals emitted by the tumor microenvironment, which hinder CAR T-cells' survival and activity. Together, these processes undermine the therapeutic effectiveness of CAR T-cell therapy over time. 

An overview of the parameters, their units, biological interpretation, as well as estimated values used in Section \ref{sec:results_math} are given in Table \ref{tab:model_full}.

The model assumes a pronounced CAR T-cell activity at the tumor's surface. This dependency is governed by the sphericity parameter $\spher$, which adjusts for the tumor's shape and structural irregularities, impacting its accessibility to CAR T-cells. Particularly, this term is incorporated in the saturation term of the CAR T-cell stimulation, emphasizing how tumor geometry influences CAR T-cell proliferation and its constraints. CAR T-cell proliferation is driven by the interaction with antigen-presenting tumor cells, capturing the process of clonal expansion triggered by antigen recognition. This expansion is a key mechanism by which CAR T-cells amplify their cytotoxic response to lymphoma B-cells. The numerator in the proliferation term quantifies the strength of this stimulation, governed by a parameter representing CAR T cells' intrinsic capacity to expand upon encountering tumor antigens. The proliferation term includes a saturation factor in the denominator, accounting for the limitations on CAR T-cell expansion in the tumor microenvironment. At low tumor burdens, CAR T-cell proliferation scales proportionally with tumor size, as antigen presentation and resources are sufficient to support expansion. However, at higher tumor burdens, factors such as limited antigen presentation, resource competition, and increased immunosuppressive signals attenuate the proliferation rate. 
Finally, the equation includes a term describing the natural decay of CAR T-cells, characterized by a decay rate $m$.

\begin{table}[h!]
\caption{Description of parameters in Equation \eqref{eq:model_full_1a} and \eqref{eq:model_full_1b}, their units, as well as the estimated parameter values.}
\centering
\begin{tabularx}{\textwidth}{ c|c|X|c|c }
\hline
Parameter & Unit & Description & Value & Source  \\
\hline
$a$ & day$^{-1}$  & tumor growth rate &  $[0.01,0.5]$ & parameter estimation\\ 
$b$ & cells$^{-1}$ & inverse of the carrying capacity of tumor cell population & $1 \times 10^{-13}$ & \cite{roeschModellingLymphomaTherapy2014} \& Section \ref{sec:methods_par_estimate}   \\ 
$d$ & day$^{-1}$ cells$^{-1}$ & tumor cell elimination rate caused by CAR T-cells & $1.101 \times 10^{-7}$ & \cite{roeschModellingLymphomaTherapy2014}\\ 
$j$ & day$^{-1}$ & maximal recruitment rate of CAR T-cells (stimulated by tumor cells) & $0.231$ & parameter estimation \\
$g$ & cells & number of tumor cells at which the stimulated recruitment rate reaches its half-maximum (smaller $g$ implies a faster increase of CAR T-cell stimulation) & $\left(2.019 \times 10^{5} \right)^{3/2}$ & \cite{roeschModellingLymphomaTherapy2014}\\ 
$q$ & day$^{-1}$ cells$^{-1}$ & CAR T-cell inactivation rate caused by tumor cells & $4.424\times 10^{-9}$ & parameter estimation\\ 
$m$ & day$^{-1}$ & CAR T-cell inactivation rate independent of tumor cells & $0.007 $ & \cite{roeschModellingLymphomaTherapy2014}\\
$\Psi$ &  & tumor lesion sphericity & $(0, 1]$  & baseline PET scans \\
$B_0$ & cells & initial number of cancerous B-cells used for model simulations & $6.598 \times 10^9$ & baseline PET scans \\
$C_0$ & cells & initial number of CAR T-cells used for model simulations & $2.020 \times 10^8$ & baseline PET scans \\
\hline
\end{tabularx}
\label{tab:model_full}
\end{table} 

The equilibria of the system given by Equation \eqref{eq:model_full_1a} and \eqref{eq:model_full_1b}
are found by computing the intersections of the nullclines $dB/dt = 0$ and $dC/dt = 0$ (SI).  In the case of $\Psi = 1$, the system establishes four equilibria $E_0 = (\bar{B}_0, \bar{C}_0) = (0,0)$, $E_1=(\bar{B}_1, \bar{C}_1) = (5.043 \times 10^5, 1.084 \times 10^8)$ cells, $E_2=(\bar{B}_2, \bar{C}_2) = (3.581 \times 10^{11}, 9.328 \times 10^9)$ cells, and $E_3 = (\bar{B}_3, \bar{C}_3) = (b^{-1}, 0)$. 
We note that by decreasing the sphericity $\Psi$, it is possible for the system to transit into another qualitative setting. However, for the chosen set of parameters, this does not occur for $\Psi \geq 0.162$, which is the minimum sphericity we observe in the clinical data used in this study. 

\subsection{Model parameter estimation}\label{sec:methods_par_estimate}
We used the total peripheral CAR T-cell concentration from the peripheral blood measurements, recorded at 7, 14, 21, and 28 days post-CAR T-cell infusion of the 63 patients. The values were given in $\mathrm{cells/\mu l}$. The total number of CAR T-cells was estimated by assuming an equal distribution of CAR T-cells in the blood for all patients. For a male adult, we assumed a blood volume of $5.5 \, \mathrm{l}$, and for a female adult, this was assumed to be $4.5 \, \mathrm{l}$. The initial CAR T-cell dose was known only for patients receiving the treatment tisa-cel. To estimate the missing values, we used the MATLAB function \texttt{fillgaps()} using autoregressive modeling on the square root of the available data to ensure non negative imputed initial CAR T-cell dose. 
Missing values in the CAR T-cell data post-infusion were handled analogously for each patient with \texttt{maxlen} set to three. Initial tumor burden was estimated from the total metabolic tumor volume, assuming a malignant B-cell had a diameter of approximately $20 \, \mathrm{\mu m}$. In accordance with \cite{roeschModellingLymphomaTherapy2014}, we fixed the tumor cell elimination rate mediated by CAR T-cells $d$ to $1.101 \times 10^{-7}$, the number of tumor cells at which the stimulated recruitment rate reaches its half-maximum $g$ to $(2.019 \times 10^5)^{3/2}$, and the tumor-independent CAR T-cell inactivation rate $m$ to $0.007$.

For the carrying capacity of the tumor cell population, we noted that the number of cells in a human body is on the order of $10^{14}$ \cite{albertsMolecularBiologyCell2002}. In their model simulations of lymphoma therapy and outcomes, \citet{roeschModellingLymphomaTherapy2014} used $10^{13}$ as the threshold for the number of tumor cells at which the simulation was stopped. Considering that we were modeling only a single lymphoma lesion and that the estimated maximum number of initial tumor cells in our data was $6.045 \times 10^{11}$, we assumed that the tumor burden could not exceed a size of $10^{13}$.

We fitted the data to the real part of the model solution $C(t)$ using the default settings of the MATLAB function \texttt{lsqnonlin()} to obtain the optimal parameters for $j$, $q$, $a$, and $\spher$, which minimized our predefined loss function. This was given by a weighted-least squares function, where the weights were the variance of the data points at each time point. The initial guess, lower, and upper bounds for the parameters are provided in Table \ref{tab:parameter_bounds}.

\begin{table}[h!]
\caption{Initial guess, lower, and upper bound of the parameters to fit.}
\centering
\begin{tabular}{ c|c|c|c|c }
\hline
Parameter & Unit & Initial guess & Lower bound & Upper bound  \\
\hline
$a$ & day$^{-1}$  & 0.18 & $0.01$ & 0.5 \\ 
$j$ & day$^{-1}$ & 0.3 & 0.01 & $\infty$ \\
$q$ & day$^{-1}$ cells$^{-1}$ & $3 \times 10^{-9}$ & 0 & $\infty$\\ 
$\spher$ &  & 1.25 & 1  & $\infty$\\
\hline
\end{tabular}
\label{tab:parameter_bounds}
\end{table}

For each of the 63 patients, the parameters were estimated by minimizing the previously defined loss function. Table \ref{tab:parameter_estimates} gives an overview of the 0.25, 0.5, and 0.75 quantile of the parameter estimates.

\begin{table}[h!]
\caption{0.25, 0.5, and 0.75 quantile of the estimated parameter values.}
\centering
\begin{tabular}{ c|c|c|c|c|c|c }
\hline
Parameter & Minimum & 0.25 quantile & Median & Mean & 0.75 quantile & Maximum  \\
\hline
$a$ & 0.010  & 0.116 & $0.186$ & 0.207 & 0.279 &0.500\\ 
$j$ & 0.010 & 0.130 & 0.231 & 0.274& $0.307$ &1.209 \\
$q$ & $3.844 \times 10^{-11}$ & $2.148 \times 10^{-9}$ & $4.424 \times 10^{-9}$ & $2.434 \times 10^{-8}$ & $1.339 \times 10^{-8}$ & $2.418 \times 10^{-7}$\\ 
$\spher$ & & 1.412  & 5.668 & 33.383 & 21.513 & 425.515\\
\hline
\end{tabular}
\label{tab:parameter_estimates}
\end{table} 
 
\backmatter

\bmhead{Supplementary information}
Supplementary information is provided in a separate file.

\bmhead{Data availability}
The data that support the findings of this study is contained in the supplementary information. 

\bmhead{Author contributions}
YC: Data processing (feature extraction), data analysis (initial data exploration and analysis), model analysis (development, analysis, and numerical simulations), writing (original draft, final draft, review, and editing). SS: Project development (initialization and coordination), data collection, data processing (feature extraction), data analysis (initial exploration and analysis), model analysis (numerical simulations), writing (original draft, final draft, review, and editing). CK: Model analysis (development and analysis), writing (review and editing). JBB: Writing (review and editing).
AMR: Data analysis (initial data exploration), writing (review and editing). VMPG: Conceptualization, project development (initialization and coordination), data collection, writing (original draft, final draft, review, and editing). LMM, LLC: Data collection and analysis (CAR T-cell quantification), writing (review and editing). JCC, CMF, MPTA: Data collection and processing (PET image acquisition and segmentation), writing (review and editing). AC, PB, PB, APM: Conceptualization, project initialization and development, writing (review and editing).

\bmhead{Acknowledgements}
This project has been funded by an investigational grant from Novartis Spain and GETH-TC and was partially supported by grant PID2022-142341OB-I00, funded by
Ministerio de Ciencia e Innovación/Agencia Estatal de Investigación, Spain (doi:10.13039/501100011033) and European Regional Development Fund (ERDF A way
of making Europe), grants SBPLY/21/180501/000145 and SBPLY/23/180225/000041 funded by Junta de Comunidades de Castilla-La Mancha. Spain (and European
Regional Development Fund (and European Regional Development Fund ERDF A way of making Europe); University of Castilla-La Mancha / ERDF, A way of making
Europe (Applied Research Projects) under grant 2022-GRIN-34405.



\bibliographystyle{abbrv}
\bibliography{bibliography}

\end{document}